\documentclass{PoS}

\usepackage{amssymb,fontenc,times,mathptmx,graphicx}
\usepackage{amsmath,latexsym}
\usepackage{axodraw,mathrsfs}
\usepackage{bbm}
\usepackage{macros}
\usepackage{psfig,tables}
\usepackage[dvips]{epsfig}
\usepackage[dvips]{graphicx}

\newcommand{\Ref}[1]{Ref.~#1}
\newcommand{\Tab}[1]{Tab.~#1}
\newcommand{\Refs}[1]{Refs.~#1}
\newcommand{\Fig}[1]{Fig.~#1}

\newcommand{\ssf}{sSF}
\newcommand{\csf}{$\chi$SF}
\newcommand{\zf}{z_{\mathrm{f}}}
\newcommand{\zfc}{z_{\mathrm{f}}^\ast}
\newcommand{\ds}{d_\mathrm{s}}
\newcommand{\kap}{\kappa}
\newcommand{\kapc}{\kappa_{\mathrm{cr}}}
\newcommand{\mpcac}{m_\mathrm{PCAC}}
\newcommand{\gam}{g_{\mathrm{A-}}}
\newcommand{\oaimp}{$O(a)$ improvement}

\title{Chirally rotated Schr\"odinger functional:\
non-perturbative tuning in the quenched approximation}

\ShortTitle{Chirally rotated Schrodinger functional}

\author{
\vspace*{-10mm}

\begin{flushright}
HU-EP-09/43\\
SFB/CPP-09-85\\
DESY 09-146\\
LTH 841\\
\end{flushright}
\vspace*{7mm}

\speaker{J.~Gonzalez Lopez}~$^{a,b}$,
K.~Jansen$^b$, D.~B.~Renner$^b$ and A.~Shindler$^c$\footnote{Current address: Instituto de F\'{\i}sica Te\'orica UAM/CSIC
Universidad Aut\'onoma de Madrid, Cantoblanco E-28049 Madrid, Spain}\\
\llap{$^a$}Humboldt-Universit\"at zu Berlin, Institut f\"ur Physik\\
Newtonstrasse 15, 12489 Berlin, Germany\\
\llap{$^b$}NIC, DESY\\
Platanenallee 6, 15738 Zeuthen, Germany\\
\llap{$^c$}Division of Theoretical Physics, University of Liverpool\\
Peach Street, Liverpool L69 7ZL, United Kingdom \\
E-mail:\ \email{jenifer.gonzalez.lopez@desy.de}
}

\abstract{
The use of chirally rotated boundary conditions provides a formulation
of the Schr\"odinger functional that is compatible with automatic {\oaimp}
of Wilson fermions in the bulk.  The elimination of bulk $O(a)$ terms requires
the non-perturbative tuning of the critical mass and one additional 
boundary counterterm.
We present the results of such a 
tuning in the quenched approximation at three values of the
renormalised gauge coupling and for a range of lattice spacings.
}

\FullConference{The XXVII International Symposium on Lattice Field Theory\\
		 July 26-31, 2009\\
		 Peking University, Beijing, China}

\begin{document}

\section{Introduction}
\label{sec:chiSF}

Obtaining precise physical results from lattice calculations requires
a well controlled continuum limit and, for many quantities,
non-perturbative renormalisation. Ideally the renormalisation scheme should
be not only \emph{non-perturbative} but also \emph{mass independent} and preferably
\emph{gauge invariant}. Schr\"odinger functional (SF)
schemes~\cite{luescher1,sint1,luescher2} are known to fulfill these
properties.  Additionally, to ease the burden of taking the continuum
limit, {\oaimp} is highly desirable.  However, to eliminate the many
counterterms necessary when applying the standard {\oaimp} program with Wilson fermions, 
we would like to capitalize on the automatic {\oaimp} provided by maximally twisted mass
fermions~\cite{FR1} (see \cite{andrea} for a review).  
Unfortunately, bulk automatic {\oaimp} with Wilson fermions
and the standard SF (sSF) boundary conditions (BCs) are not compatible.
{\oaimp} is only possible introducing a number of additional
bulk improvement counter-terms to the action and operators.
Since there are extensive calculations with maximally twisted mass
fermions \cite{Boucaud:2007uk,Boucaud:2008xu} it would be clearly
desirable to employ the SF scheme while keeping
automatic $O(a)$-improvement.

A new formulation of the SF has been developed in \Ref{\cite{sint2}},
which we will refer to as the {\emph chirally rotated} SF ({\csf}),
that implements a SF scheme while maintaining automatic {\oaimp} for massless Wilson fermions.
The {\csf} is related (in the continuum) to the {\ssf}
by means of a non-singlet chiral transformation, i.e. they are equivalent in the continuum limit.
However, when using massless Wilson fermions as a lattice regulator,
{\csf} BCs are invariant under a subgroup of the chiral symmetry 
transformations broken by the Wilson term (in contrast
to {\ssf} BCs). As a result {\csf} BCs are compatible with automatic {\oaimp}.

The three-dimensional boundaries of the SF lead to an unavoidable
dimension four boundary operator.  Additionally, regulating the {\csf}
with Wilson fermions induces the usual bulk mass operator as 
well as a dimension three boundary operator.  The
dimension four boundary operator is irrelevant, and hence the
corresponding coefficient can be safely fixed by perturbation theory
in order to eliminate the corresponding $O(a)$ boundary contributions.
The bulk operator is relevant and is handled by the
standard non-perturbative tuning of the bare quark mass, equivalently $\kap$, to its critical value.
The dimension three operator is also relevant and can spoil not only the
automatic {\oaimp} but also the universality of the continuum limit.
This requires an additional non-perturbative tuning of one more
counterterm, $\zf$.  However, having tuned both $\kap$ and $\zf$,
all operators are automatically $O(a)$ improved and no further counterterms are
necessary.

Here we present the non-perturbative tuning of $\kap$ and $\zf$ for
the {\csf} in the quenched approximation.  We demonstrate the
feasibility of tuning both parameters simultaneously.  In particular,
the inclusion of the bulk dimension five operator, with corresponding counterterm
$c_{\mathrm{sw}}$, as used in \Ref{\cite{sintleder}}, is found to be
unnecessary.

\section{Boundary conditions}

The {\csf} is related to the {\ssf} by a non-singlet chiral
transformation, $\chi = \exp(-i\pi \gamma_5 \tau^3/4)\psi$, where
$\psi$ is the fermion doublet in the $N_{\rm f} = 2$ standard formulation, $\chi$
is the corresponding doublet in the rotated basis and $\tau^3$ is a Pauli matrix. 
This field transformation maps the {\ssf} BCs to the {\csf} BCs,
\begin{align}
\label{eq:contbc}
Q_{+}\chi(x)|_{x_{0} = 0}& =0&
Q_{-}\chi(x)|_{x_{0} = T}& =0\\
\overline{\chi}(x)Q_{+}|_{x_{0} = 0}& =0&
\overline{\chi}(x)Q_{-}|_{x_{0} = T}& =0\,, \nonumber
\end{align}
where $T$ is the Euclidean time extent and $Q_{\pm}$ are projectors
given by
\begin{displaymath}
Q_{\pm} = \frac{1}{2}\, \left( \mathbbm{1} \pm i\, \gamma_{0}\gamma_{5}\tau^{3} \right)\,.
\end{displaymath}
Thus the $Q_{\pm}$ are simply the chirally rotated projectors
corresponding to the {\ssf} projectors, $P_{\pm} = 1/2(1\pm\gamma_0)$.
However, once the theory is regularised on the lattice, we must ensure
that the BCs in \eqref{eq:contbc} are in fact recovered in the
continuum limit.  Using orbifolding techniques, it was shown that the
BCs can be implemented at finite lattice spacing by a simple
modification of the standard Wilson-Dirac operator, ${D}_{\rm W}$,
near the time boundaries~\cite{sint4}.  The resulting action is
\begin{equation}
\label{eq:action}
S = a^4 \sum_{x_0=0}^{T}\sum_{\rm{\vec{x}}}\overline\chi(x)\left({\mathcal D}_{\rm W} + m_0\right)\chi(x)
\end{equation}
and the modified Wilson-Dirac operator is given by
\begin{equation}
a {\mathcal D}_{\rm W}\chi(x) = \left\{ 
  \begin{array}{ l l }
     -U(x,0)P_-\chi(x+a\hat{0}) + (aK +i \gamma_5\tau^3 P_- )\chi(x) & \qquad {\rm if} \quad x_0=0 \\
     a {D}_{\rm W}\chi(x) & \qquad {\rm if} \quad 0 < x_0 < T \\
     (aK +i \gamma_5\tau^3 P_+ )\chi(x)-U^{\dagger}(x-a\hat{0},0)P_+\chi(x-a\hat{0}) & \qquad {\rm if} \quad x_0=T \\
  \end{array} \right.
\label{eq:latact}
\end{equation}
where $K$ is the time-diagonal contribution to ${D}_{\rm W}$.

\section{Boundary counterterms}

To ensure the correct continuum limit, we must account for all
relevant operators allowed by the symmetries of the action above.
This means dimension four or less for the bulk action. There is one
such operator, $\overline{\chi}\chi$, and the corresponding counterterm is the 
term proportional to the critical quark mass, $m_{\rm cr}$, or equivalently $\kappa_{\rm cr}$.
This is the
standard operator that is present for all Wilson actions due to the
breaking of chiral symmetry by the Wilson term.

Similarly, we must include all permitted boundary operators of
dimension three or less.  Again, the one allowed operator is
$\overline{\chi}\chi$~\cite{sint2}, which gives rise to the following
counterterm to the lattice action,
\begin{displaymath}
\delta S_3 = 
(\zf - 1) a^{3}\sum_{\vec{x}}\, \left( \overline{\chi}\chi|_{x_{0}=0} + \overline{\chi}\chi|_{x_{0}=T} \right)\,.
\end{displaymath}
Such an operator would be forbidden in the continuum action, but the reduced symmetries of the
Wilson action do not allow us to exclude this operator on the lattice.
The presence of
$\delta S_3$ can then be understood as necessary to restore the symmetries 
broken by the Wilson term in the continuum limit.  The fact that it is a relevant
operator implies that we must compute the bare coupling
dependence of $\zf$ non-perturbatively, just as for $\kap$.

Furthermore, we must examine those irrelevant operators that lead to
$O(a)$ contributions.  In the bulk, there is the dimension five
Sheikholeslami-Wohlert term, but automatic {\oaimp} eliminates the
need for this operator.
Yet, there does remain an $O(a)$
contribution from the boundary due to the irrelevant dimension four
operator~\cite{sintleder},
\begin{displaymath}
  \delta S_4 = (\ds - 1) a^{4}\sum_{\vec{x}}\,
  \left( \overline{\chi}\gamma_{k}D_{k}\chi|_{x_{0}=0} + \overline{\chi}\gamma_{k}D_{k}\chi|_{x_{0}=T} \right).
\end{displaymath}
Such a contribution is present in all SF
formulations~\cite{luescher2} and is not due to the particular lattice
action or BCs we have chosen.  In fact, $\ds$ plays a role that is
analogous to the $\tilde{c}_t$ counterterm in the
{\ssf}~\cite{ct_tilde}.  Given that $\delta S_4$ is an irrelevant
operator, $d_s$ can be computed in perturbation theory.  For the
investigation presented here, we simply use the tree-level value of
$1/2$.

\section{Tuning conditions}

The non-perturbative determination of $\kap$ and $\zf$ requires
imposing conditions at finite lattice spacing that ensure the
restoration of all expected symmetries in the continuum limit: 
parity and flavour symmetries in the $\chi-$ basis\footnote{We recall that in the $\chi-$
basis parity and flavour symmetries take a slightly different form (see ref.~\cite{andrea} for a discussion
about the dependence of the symmetries on the basis adopted).}.  
Moreover, these conditions should be imposed at each lattice spacing
while fixing a suitable renormalised quantity. In this work, we keep
the renormalised SF coupling, $\overline{g}$, fixed.  This is
equivalent to fixing the physical size of the box, $L$.
All other dimensionful quantities must scale with $L$, so we
choose $T=L$, evaluate all correlation functions at $x_0 = T/2$ and
use periodic boundary conditions with $\theta=0$.

Before specifying the tuning conditions, we define the following
boundary to bulk correlation functions
\begin{displaymath}
g_{\mathrm{A}_{\pm}}^{ab}(x_{0}) = -\langle A_{0}^{a}(x)\mathcal{Q}_{\pm}^{b}\rangle
\qquad
g_{\mathrm{P}_{\pm}}^{ab}(x_{0}) = -\langle P^{a}(x)\mathcal{Q}_{\pm}^{b}\rangle
\end{displaymath}
where the boundary operator, $\mathcal{Q}_{\pm}^{a}$, is defined for the $x_0=0$ boundary by
\begin{displaymath}
\mathcal{Q}_{\pm}^{a} = a^{6}\sum_{{\vec{y}},{\vec{z}}} 
\overline{\zeta}({\vec{y}})\gamma_{5}\frac{1}{2}\tau^{a}Q_{\pm}\zeta({\vec{z}})\, e^{i{\vec{p}}({\vec{y}}-{\vec{z}})}\,,
\end{displaymath}
the bulk operators $A_{\mu}^{a}(x)$ and $P^{a}(x)$ are the axial
current and pseudoscalar density in the $\chi$-basis, and the boundary
fields for $x_0=0$ are defined as
\begin{displaymath}
\zeta({\vec{x}}) = U(x_{0}-a,{\vec{x}};0)\chi(x)|_{x_{0}=a}
\qquad
\overline{\zeta}({\vec{x}}) = \overline{\chi}(x)U^{\dagger}(x_{0}-a,{\vec{x}};0)|_{x_{0}=a}.
\end{displaymath}

To tune $\kap$ to its critical value, we adopt the standard procedure
of imposing a vanishing PCAC mass. To tune $\zf$, we require the
$\gamma_{5}\tau_{1}$-odd correlation function
$g_{\mathrm{A}_{-}}^{11}$ to vanish,
\begin{equation} 
\mpcac \equiv \frac{\partial_{0}^{\mathrm{latt}} g_{\mathrm{A}_{-}}^{11}(T/2)}{2g_{\mathrm{P}_{-}}^{11}(T/2)} = 0
\qquad
\gam \equiv g_{\mathrm{A}_{-}}^{11}(T/2) = 0\,.
\end{equation}
The second condition in particular is sensitive to the symmetries broken by the lattice action
\eqref{eq:action},
and both conditions together ensure that in the continuum limit all
broken symmetries are indeed restored.  Imposing different symmetry
restoration conditions would give rise to different values of $\kap$
and $\zf$ that would differ amongst themselves by cutoff effects. It
will be important to study the sensitivity of $\kap$ and $\zf$ to the
particular definitions used in order to better understand the
intrinsic uncertainty in the determination of these counterterms.

\section{Tuning results}

To check the practicality of tuning both $\kap$ and $\zf$
non-perturbatively for the {\csf}, we perform the tuning at three
values of the renormalisation scale $\mu=1/L$, corresponding to a
hadronic ($\overline{g}^{2}$ fixed with $L = 1.436r_{0}$), an
intermediate ($\overline{g}^{2}=2.4484$) and a perturbative
($\overline{g}^{2}=0.9944$) scale.  The results at these three points
are summarised in \Tab{\ref{tab:tuning}}.
\begin{table}
\begin{center}
\begin{tabular}[c]{|r|l|l|l|l|}\hline
\multicolumn{1}{|c|}{$L/a$} &
\multicolumn{1}{|c|}{$\beta$} &
\multicolumn{1}{|c|}{$\zfc$ ({\csf})} &
\multicolumn{1}{|c|}{$\kapc$ ({\csf})} &
\multicolumn{1}{|c|}{$\kapc$ ({\ssf})} \\\hline
\multicolumn{5}{|c|}{Tuning at a hadronic scale, $\mu \sim 300\textrm{ MeV}$} \\\hline
 8 & 6.0219 & 1.8090\,(32) & 0.153530\,(24) & 0.153371\,(10) \\
10 & 6.1628 & 1.7920\,(30) & 0.152134\,(17) & 0.152012\,(7)  \\
12 & 6.2885 & 1.7664\,(51) & 0.150815\,(22) & 0.150752\,(10) \\
16 & 6.4956 & 1.7212\,(83) & 0.148945\,(25) & 0.148876\,(13) \\\hline
\multicolumn{5}{|c|}{Tuning at an intermediate scale, $\mu \sim 1\textrm{ GeV}$} \\\hline
 8 & 7.0197 & 1.5467\,(15) & 0.144501\,(13) & 0.144454\,(7) \\
12 & 7.3551 & 1.5126\,(23) & 0.143113\,(12) & 0.143113\,(6) \\
16 & 7.6101 & 1.4942\,(37) & 0.142112\,(13) & 0.142107\,(6) \\\hline
\multicolumn{5}{|c|}{Tuning at a perturbative scale, $\mu \sim 30\textrm{ GeV}$} \\\hline
8  & 10.3000 & 1.29730\,(67) & 0.1354609\,(54) & 0.135457\,(5) \\
12 & 10.6086 & 1.2954\,(11)  & 0.1351758\,(56) & 0.135160\,(4) \\
16 & 10.8910 & 1.2858\,(15)  & 0.1348440\,(61) & 0.134849\,(6) \\\hline
\end{tabular}
\caption{Tuning results at a hadronic, intermediate and perturbative
scale.  We give the critical values, $\zfc$ and $\kapc$, calculated
in this work for the {\csf}.  For reference, we also give $\kapc$
for the {\ssf}~\cite{ref1,ref2,ref3}.}
\label{tab:tuning}
\end{center}
\end{table}
We now briefly explain the procedure we used to perform the tuning,
showing examples from our most difficult point at the hadronic scale
and for the smallest lattice, $L/a=8$.

\begin{figure}
\begin{minipage}{214pt}
\includegraphics[width=165pt,angle=270]{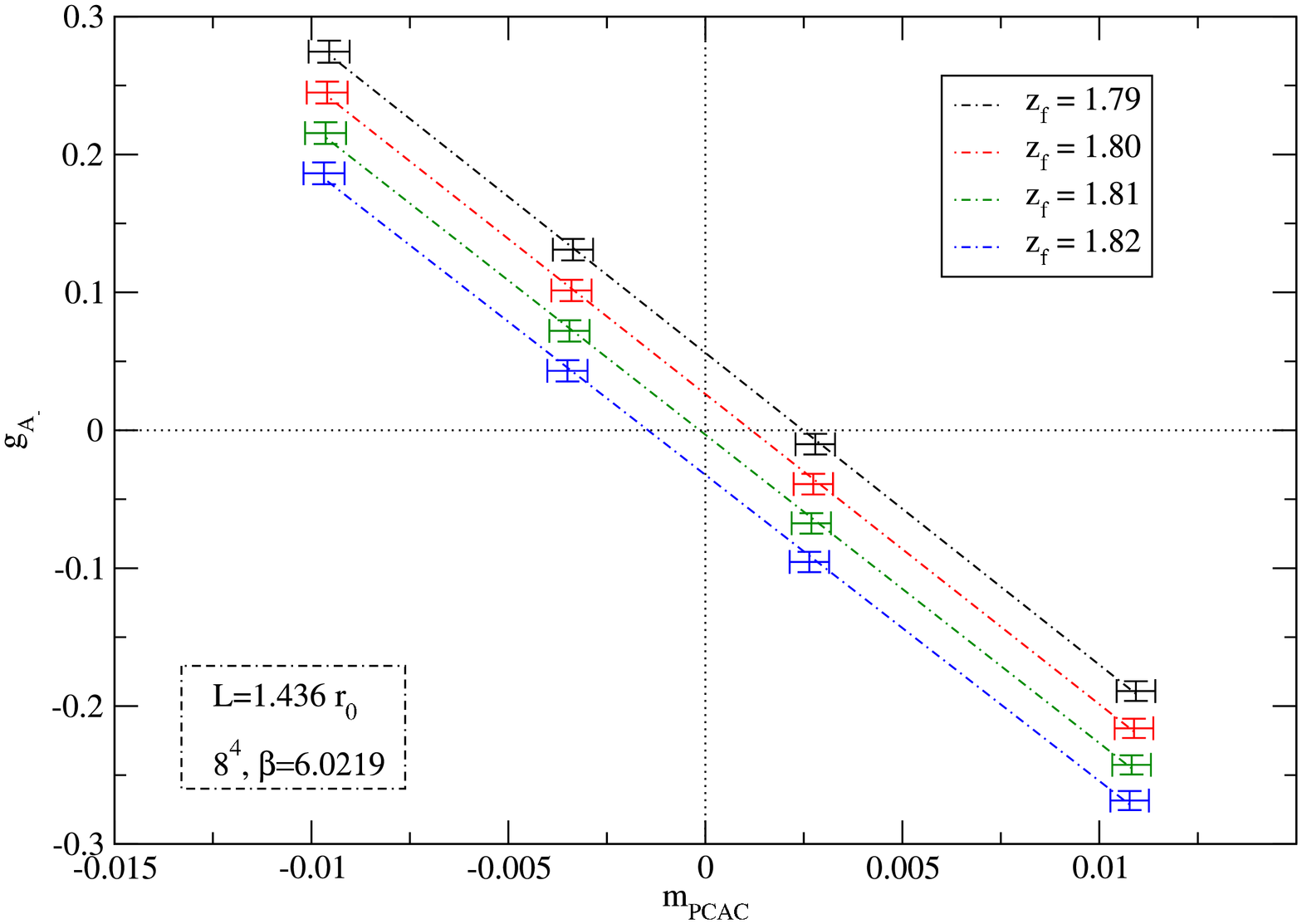}
\caption{Plot of $\gam$ versus $\mpcac$.}
\label{fig:ga_vs_mpcac}
\end{minipage}
\begin{minipage}{214pt}
\includegraphics[width=165pt,angle=270]{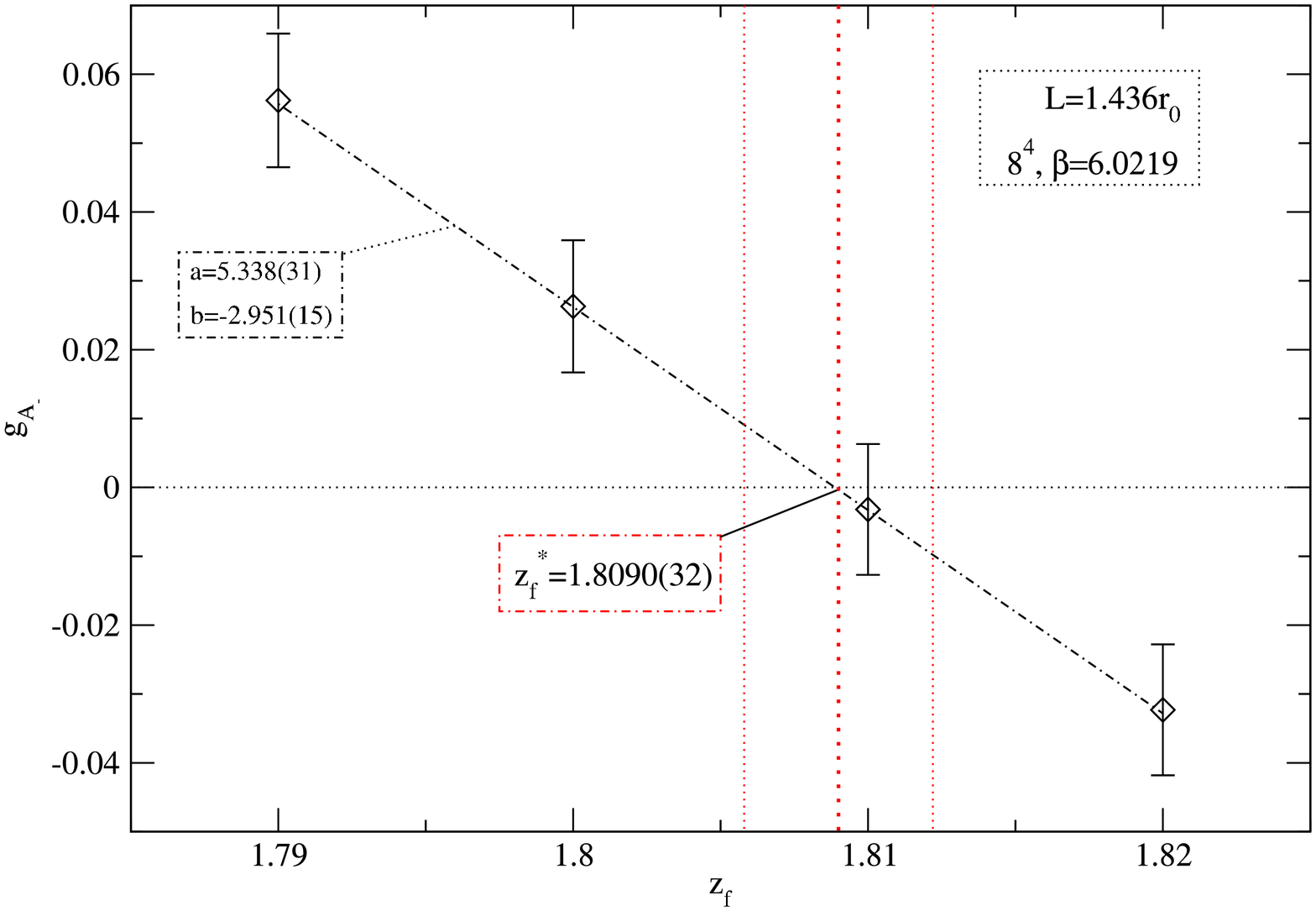}
\caption{Plot of $\gam$ versus $\zf$.
\label{fig:ga_vs_zf}}
\end{minipage}
\end{figure}

The values of $\beta$ used are given in \Tab{\ref{tab:tuning}} and are
taken from \Ref{\cite{alpha1}}.  The tuning is performed in several
steps.  First, we calculate $\mpcac$ and $\gam$ at four values of
$\zf$, and for each value of $\zf$, we use four values of $\kap$, thus
giving 16 pairs of $\kap$ and $\zf$.  This allows us to determine
$g_{A_{-}}$ as a function of $\mpcac$ for each value of $\zf$, as
illustrated in \Fig{\ref{fig:ga_vs_mpcac}}.
For each value of $\zf$, we perform a linear interpolation of $\gam$
in terms of $\mpcac$ to the point $\mpcac=0$.  This determines the
values of $\gam$ at $\mpcac=0$ for each of the four values of $\zf$,
as shown in \Fig{\ref{fig:ga_vs_zf}}.  We now interpolate these values
of $\gam$ as a function of $\zf$ to the point of vanishing $g_{A_{-}}$, thus giving us the critical value $\zfc$.

\begin{figure}
\begin{minipage}{214pt}
\includegraphics[width=165pt,angle=270]{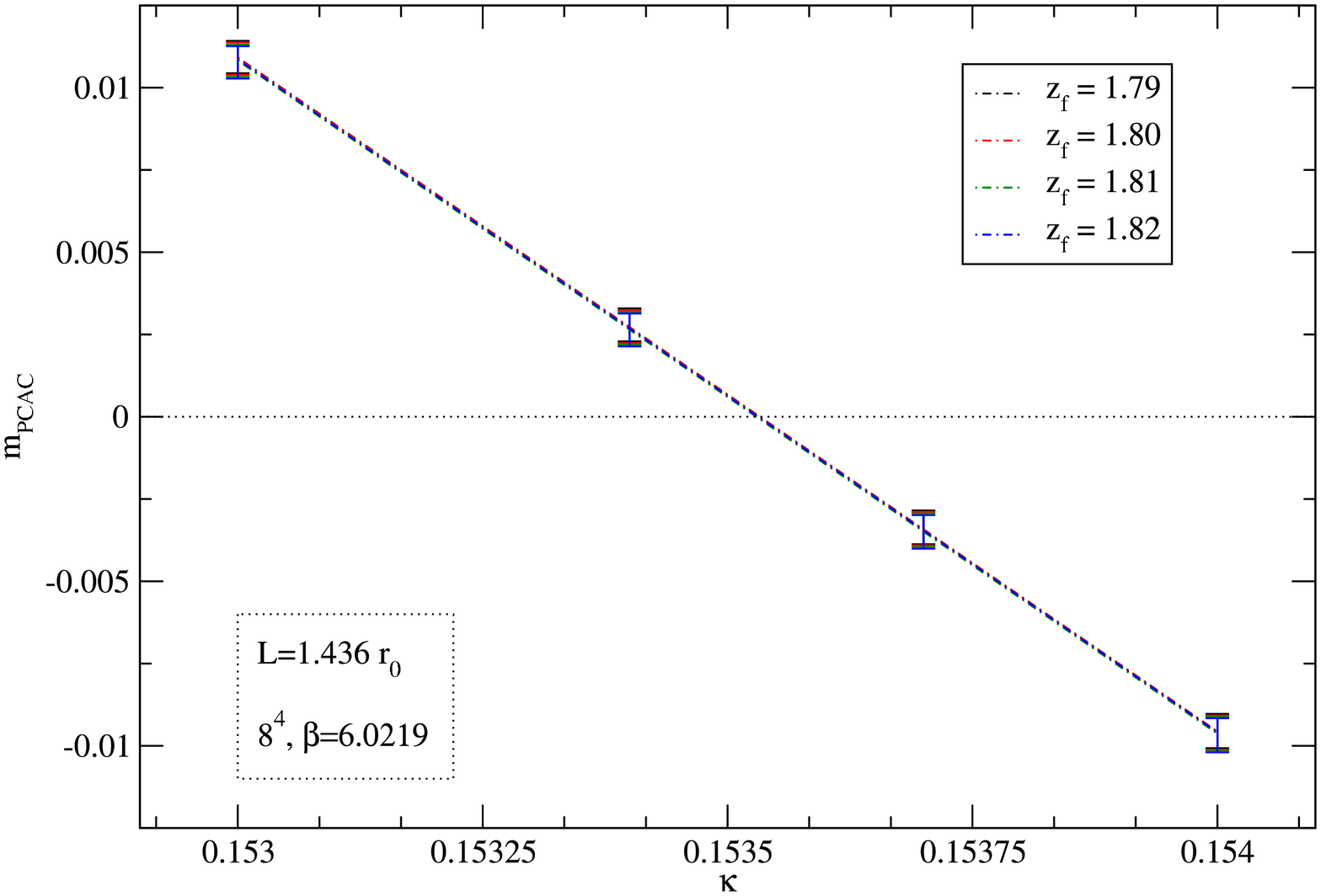}
\caption{Plot of $\mpcac$ versus $\kap$.
\label{fig:mpcac_vs_kappa}}
\end{minipage}
\begin{minipage}{214pt}
\includegraphics[width=165pt,angle=270]{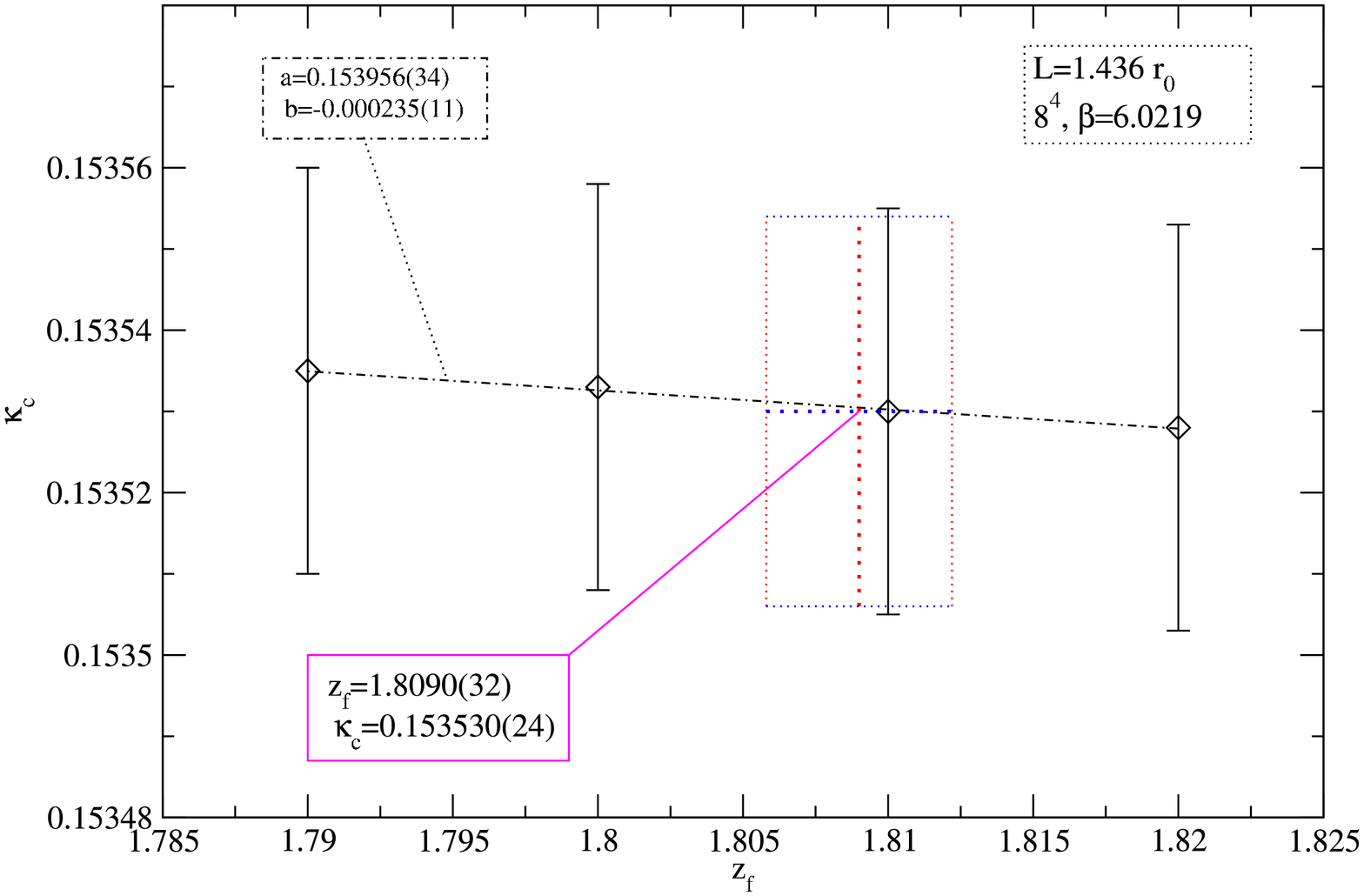}
\caption{Plot of $\kapc$ versus $\zf$.
\label{fig:kappac_vs_zf}}
\end{minipage}
\end{figure}

Next we determine $\kapc$.  Using the same 16 pairs of $\kap$ and
$\zf$, we calculate $\mpcac$ as a function of $\kap$ for each $\zf$.
This is shown in \Fig{\ref{fig:mpcac_vs_kappa}}.  Note that $\mpcac$
has a very mild dependence on $\zf$, so the four curves at fixed $\zf$
are nearly indistinguishable.  Interpolating in $\kap$ to the point of
vanishing PCAC mass, we obtain the critical values of $\kap$ at each
$\zf$.  The resulting values of $\kap$ as a function of $\zf$ are
shown in \Fig{\ref{fig:kappac_vs_zf}}.
We now interpolate these results in $\zf$ to the previously determined
value of $\zfc$, thus determining the value of $\kapc$.

A key observation of this work is the mild dependence of $\mpcac$ on
$\zf$, at least in the region near $\kapc$ and $\zfc$.  You can easily see
this in \Fig{\ref{fig:mpcac_vs_kappa}}.  The consequence of
this is clear in \Fig{\ref{fig:kappac_vs_zf}}:\ the determination of
$\kapc$ also has a weak dependence on $\zfc$ and the errors of both
are relatively independent.  If this behaviour persists with dynamical
calculations, it could ease the numerical effort necessary to perform
the tuning, thus reducing the number of required simulations.

\section{Conclusions}

We have presented the results of the non-perturbative tuning of $\kap$
and $\zf$ for the {\csf} at three physical scales and for a range of
lattice spacings.  This demonstrates that the tuning of these two
coefficients is indeed feasible, at least in the quenched
approximation.  Moreover, we observe that the tuning of $\zf$ and
$\kap$ are nearly independent.  Note that even with non-improved
Wilson fermions in the bulk, $\kap$ and $\zf$ are the only parameters
that must be tuned within the {\csf} setup in order to guarantee bulk
automatic {\oaimp}, thus eliminating the need for the bulk
counterterm, $c_{\mathrm{sw}}$, and for the many operator improvement
coefficients necessary in the {\ssf}.

Our next step is to perform an universality test of this formulation
as well as a demonstration that automatic {\oaimp} holds.  This can be
done by reproducing a variety of quantities already computed in the
standard setup.  A natural candidate would be the computation of the
step-scaling function of the pseudoscalar renormalisation factor,
$Z_\mathrm{P}$, which could be compared to the results
of~\cite{alpha1}.  We recall that the {\csf} and the {\ssf} are
equivalent in the continuum limit, therefore, it is not necessary to
recompute the entire evolution of an operator.  The only quantity that
must be recomputed is the renormalisation factor at the most
non-perturbative scale.

We also plan to explore whether the value of $\kapc$ determined from
the finite volume simulations can be used in large volume, preserving
the nice scaling behaviour obtained in
\Refs{\cite{chilf1,Dimopoulos:2008sy}}, without the need for a large
volume determination of $\kapc$.  A lattice perturbation theory
computation of $\ds$ and $\zf$ is also planned.  The final goal is to
perform dynamical simulations.

\acknowledgments

We thank S.~Sint and B.~Leder for many discussions and the private
communication of the unpublished results of ref.~\cite{sint4}.  We also
acknowledge the support of the computer center in DESY-Zeuthen and the
NW-grid in Lancaster.

\end{document}